\documentclass[prb,aip,apl,reprint,twocolumn]{revtex4-1}
\usepackage{lipsum}
\usepackage{graphicx} 
\usepackage{bm} 
\usepackage{dcolumn}
\usepackage{amssymb,amsmath}
\bibstyle{apsrev}
\usepackage{hyperref}
\usepackage{color}
\usepackage[normalem]{ulem}

\begin{document}

\title {Equivalence between a topological and non-topological quantum dot - hybrid structures}

\author{A. M. Calle}
 \email{ana.callea@usm.cl}
\affiliation{Departamento de F\'isica, Universidad T\'ecnica Federico Santa Mar\'ia, Avenida Espa\~{n}a 1680, Casilla 110V, Valpara\'iso, Chile \\}
\author{M. Pacheco}
\affiliation{Departamento de F\'isica, Universidad T\'ecnica Federico Santa Mar\'ia, Avenida Espa\~{n}a 1680, Casilla 110V, Valpara\'iso, Chile \\}
\author{P. A. Orellana}
\affiliation{Departamento de F\'isica, Universidad T\'ecnica Federico Santa Mar\'ia, Avenida Espa\~{n}a 1680, Casilla 110V, Valpara\'iso, Chile \\}
\author{J. A. Ot\'alora}
\affiliation{Institute of Metallic Materials at the Leibniz Institute for Solid State and Materials Research, IFW, 01069 Dresden, Germany\\}

\date{\today}
\begin{abstract}
In this work we demonstrate an equivalence on the single-electron transport properties between systems of different nature, a topological quantum system and a non-topological one. Our results predicts that the Fano resonances obtained in a T-shaped double quantum dot system coupled to two normal leads and one superconducting lead (QD--QD--S) are identical to the obtained in a ring system composed of a quantum dot coupled to two Majorana bound states confined at the ends of a one dimensional topological superconductor nanowire (QD--MBSs). We show that the non-zero value of the Fano (anti)resonance in the conductance of the QD-MBSs systems is due to a complex Fano factor $q_{M}$, which is identical to the complex Fano factor $q_{S}$ of the QD--QD--S. The complex nature of $q_{S}$ can be understood as a sign of a phase introduced by the superconducting lead in the QD--QD--S. It is because of this phase that the equivalence between the QD--QD--S and the QD--MBSs is possible. We believe that our results can motivate further theoretical and experimental works toward the understanding of transport properties of topological quantum hybrid structures from conventional non-topological quantum systems. 
\end{abstract}
\pagestyle{plain} 
\maketitle


\textit{Introduction}. 
Majorana bound states (MBSs)\cite{Alicea, Beenakker2,Yoreg, DasSarma, Hasan, Sato} have been highlighted by their topologically protected role in quantum computing. In virtue of their nonlocal topological nature, MBSs are immune to local noise, \cite{DasSarma} which in contrast to ordinary quantum computation, their implementation as data carriers would not require of any quantum error correction. 

MBSs take place in quantum systems with strong spin-orbit coupling, superconductivity, and broken time-reversal symmetry. \cite{Yoreg, DasSarma, Fulga}
Promising platforms to observe MBSs involve topological superconductors realized in semiconductors, specifically, semiconductor nanowires with a strong spin-orbit coupling in proximity to an s-wave superconductor and subject to a magnetic field.\cite{Yoreg, DasSarma, Fulga,DasSarma2011, Cayao} 
In such systems, Majorana modes may be detected by measuring the zero-bias conductance peak (ZBCP).\cite{Mourik, Das, Finck, Rokhinson, Nichele, Flensberg} 
Nevertheless, distinguishing between MBSs and other spurious-zero energy modes is challenging, since the ZBCP can also be caused by disorder, multi-band effects, weak antilocalization, the Kondo effect and Andreev Bound States (ABSs). \cite{Hell} Currently, distinguishing MBSs from ABSs is one of the most critical challenges, which has lead recently to considerable experimental and theoretical efforts,\cite{Fingerprints, Yoreg1, Hell, DasSarma1, DasSarma2, Seridonio, Deng2018, Sau} mostly focused on quantum dots coupled to topological superconductors (QD--MBSs configurations). Indeed, evidence of their existence have been shown by probing their transport conductance spectrum,\cite{Liu2011,Cao2012, Seridonio} thermal conductance \cite{PhysRevLett.106.057001}, current noise \cite{Lu2016, Cao2012}, ac Josephson effect \cite{PhysRevB.86.140503,PhysRevLett.108.257001}, and particularly, Fano resonances \cite{Fano_Miroshnichenko2010, Domanski_2017,Xiong_2016,Zeng2017,Schuray2017Fano,Cayao,Xia2015Fano, Ueda2014Fano} 

In contrast to the aforementioned efforts to distinguish MBSs transport properties from the ABSs in topological superconductors, in this paper we demonstrate an interesting case where the quantum transport of MBSs  in a topological superconductive quantum interferometer\cite{Zeng2017} is equivalent to the quantum transport of ABSs in a conventional T-shape quantum dot system (without topology).\cite{ Domanski_2011, Domanski_2012,AMCalle}
We specifically show that the quantum transport of a system of two quantum dots coupled in T-shape to two metallic leads and one conventional superconductor lead - which we will denote as QD--QD--S or as the non-topological system - is equivalent to a topological configuration consisting of a QD coupled to two MBSs confined at the ends of a 1D topological superconductor nanowire - denoted here as QD--MBSs.\cite{Zeng2017} Moreover, besides that the Fano resonances give a clear signature of the existence of MBSs in the QD--MBSs,\cite{Zeng2017} we show the conditions to obtain the same Fano signal but in the QD--QD--S system, which is due to the ABSs. Furthermore, we show that the Fano effect of each system is determined by their own structural parameters.

\begin{figure}[t!]
\centerline{\includegraphics[scale=0.4,angle=0]{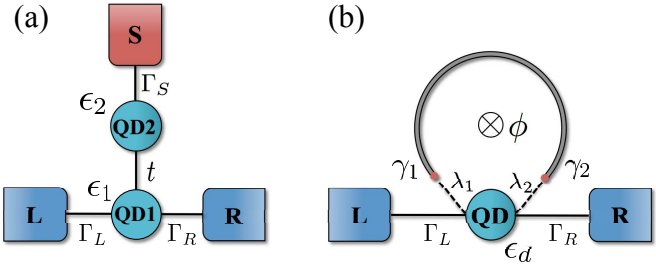}}
\caption{a) Schematic setup of the QD--QD--S system. The QDs are coupled to left ($L$) and right ($R$) normal leads and a superconductor lead ($S$) with an interdot coupling denoted by $t$. b) Schematic setup of a QD-MBSs ring.The QD is coupled to two MBS $\gamma_{1}$ and $\gamma_{2}$ with $\lambda_{1}$ and $\lambda_{2}$ respectively. $\phi$ is the phase resulting from the magnetic flux through the ring. The QD is coupled to two normal metallic leads $L$ and $R$.}
\label{system}
\end{figure}

\textit{Model}. The T-shaped double quantum dot is assumed with a single-level in each quantum dot, which are coupled to two normal metallic leads and to a superconductor lead, as shown in Fig. \ref{system} (a). The double quantum dot is modeled by a two impurity Anderson Hamiltonian and the Hamiltonian for the whole system can be written as $H=H_{L(R)}+H_{S}+H_{dot}+H_{T}$, where $H_{L(R)}$ is the Hamiltonian for the left (right) normal lead, which is given by $H_{L(R)}=\sum_{k_{L(R)}\sigma}\epsilon_{k_{L(R)}}C^{\dag}_{k_{L(R)}\sigma}C_{k_{L(R)}\sigma} \hspace{0.1cm},$ being $C^{\dag}_{k_{L(R)}\sigma}$ and $C_{k_{L(R)}\sigma}$ creation and annihilation operator for electrons with momentum $k_{L(R)}$ and spin $\sigma$ in the metallic lead $L(R)$.  The standard BCS Hamiltonian for the superconductor lead is $H_{S}=\sum_{k_{S}\sigma}\epsilon_{k_{S}}C^{\dag}_{k_{S}\sigma}C_{k_{S}\sigma}+\sum_{k_{S}}\Delta\left( C^{\dag}_{k_{S}\uparrow}C^{\dag}_{-k_{S}\downarrow}+h.c.\right)\hspace{0.1cm},$ where $C^{\dag}_{k_{S}\sigma}$ and $C_{k_{S}\sigma}$ are the creation and annihilation operators for electrons in the superconducting lead, while $\Delta$ is the superconducting gap function which is assumed to be s-wave, i.e., $k$-independent and real ($\Delta^{\dag}=\Delta$). The Hamiltonian for the double quantum dot is given by $H_{dot}=\sum_{i\sigma}\epsilon_{i}d^{\dag}_{i\sigma}d_{i\sigma}$, with $d^{\dag}_{i\sigma}$($d_{i\sigma}$) being the creation (annihilation) operator for electrons in the quantum dot level $\epsilon_{i}$ ($i=1,2$). Finally, the tunneling between the QD's and leads is described by $H_{T}=\sum_{\sigma}t\left(d^{\dag}_{1\sigma}d_{2\sigma}+d^{\dag}_{2\sigma}d_{1\sigma}\right)+\sum_{k_{S}\sigma}\left(V_{k_{S}}C^{\dag}_{k_{S}\sigma}d_{2\sigma}+h.c.\right)+\sum_{k_{L(R)}\sigma}\left(V_{k_{L(R)}}C^{\dag}_{k_{L(R)}\sigma}d_{1\sigma}+h.c.\right) \hspace{0.1cm}$. Here, the embedded quantum dot (QD1) is coupled to the side-coupled quantum dot (QD2) via the inter-dot coupling $t$, which is taken as being a real parameter. $V_{k_{L(R)}}$ and $V_{k_{S}}$ are the coupling between left (right) normal lead and QD1 and between superconducting lead and QD2, respectively.

In general, the current from the lead $\alpha$ ($\alpha$=L to $\alpha$=R) is given by $I_{\alpha}=e\langle \dot{N}_{\alpha} \rangle=\dot{\imath}\frac{e}{\hbar}\langle \left[H,N_{\alpha}\right] \rangle$, from which, we can get \cite{Jauho_book}

\begin{equation}
I_{L}=-2\frac{e}{h}\int d\epsilon \left[f_{L}\left(\omega\right)-f_{R}\left(\omega\right)\right]\frac{\Gamma_{L}\Gamma_{R}}{\Gamma_{L}+\Gamma_{R}}\Im\left[G^{r}_{1,11}\left(\omega\right)\right]\hspace{0.1cm},
\end{equation}
\noindent where $f_{L}(f_{R})$ is the fermi distribution of the left(right) metallic lead. 
Then, the zero-temperature conductance is
\begin{equation}
\label{condG}
G=-\frac{2e^{2}}{h}\frac{\Gamma_{L}\Gamma_{R}}{\Gamma_{L}+\Gamma_{R}}\Im\left[G^{r}_{1,11}\left(\omega\right)\right]|_{\omega=eV}\hspace{0.1cm}.
\end{equation}
\noindent The Green function of the embedded quantum dot $G^{r}_{1,11}\left(\omega\right)$ in the far subgap regime \cite{Domanski_2012} $|\omega|\ll \Delta$ (in which only the off-diagonal terms of the superconductor's self-energy matrix are preserved tending to the static value $\Gamma_{S}/2$) is

\begin{equation}
\label{G211}
G^{r}_{2,11}\left(\omega\right)=\frac{1}{\mathcal{D}\left(\omega\right)}\left(\left(\omega+\epsilon_{2}\right)+i\frac{\Gamma}{2}-\mathcal{N}\left(\omega\right)\left(\omega-\epsilon_{1}\right)\right)\hspace{0.1cm},
\end{equation}

\noindent where $\mathcal{D}(\omega)=\mathcal{F}_{+}(\omega)\mathcal{F}_{-}(\omega)-\left((\Gamma/2)\mathcal{N}(\omega)\right)^{2}$, $\mathcal{F}_{\pm}(\omega)=\omega\mp\epsilon_{2}+i\frac{\Gamma}{2}-\mathcal{N}\left(\omega\right)\left(\omega\pm\epsilon_{1}\right)$, and $\mathcal{N}\left(\omega\right)=t^{2}\left(\left(\omega-\epsilon_{1}\right)\left(\omega+\epsilon_{1}\right)-\left(\frac{\Gamma_{S}}{2}\right)^{2}\right)^{-1}$.

The retarded Green function of the QD for the system QD-MBSs (Fig. \ref{system} (b)) has the following form \cite{Zeng2017}

\begin{equation}
\label{G_Zeng}
G^{r}_{d,d^{\dag}}\left(\omega\right)=\left[\omega-\epsilon_{d}+i\frac{\Gamma}{2}-A\left(\omega\right)-B\left(\omega\right)\right]^{-1}\hspace{0.1cm},
\end{equation}

\noindent where $A\left(\omega\right)=K\left(|\lambda_{1}|^{2}+|\lambda_{2}|^{2}+\frac{2\epsilon_{M}}{\omega}|\lambda_{1}||\lambda_{2}|\cos\frac{\phi}{2}\right)$ and $B\left(\omega\right)=\frac{K^{2}\left(|\lambda_{1}|^{4}+|\lambda_{2}|^{4}-2|\lambda_{1}|^{2}|\lambda_{2}|^{2}\cos\phi\right)}{\left(\omega+\epsilon_{d}+i\frac{\Gamma}{2}-A\left(\omega\right)\right)}$, with $K$ and $\Gamma$ being defined as $K=\frac{\omega}{\omega^{2}-\epsilon^{2}_{M}}$ and $\Gamma=\Gamma_{L}+\Gamma_{R}$. The parameter $\epsilon_{M}$ is the overlap of the Majorana fermions $\gamma_{1}$ and $\gamma_{1}$ and $\epsilon_{d}$ is the energy of the QD.

\begin{figure}[t!]
\centerline{\includegraphics[scale=0.4,angle=0]{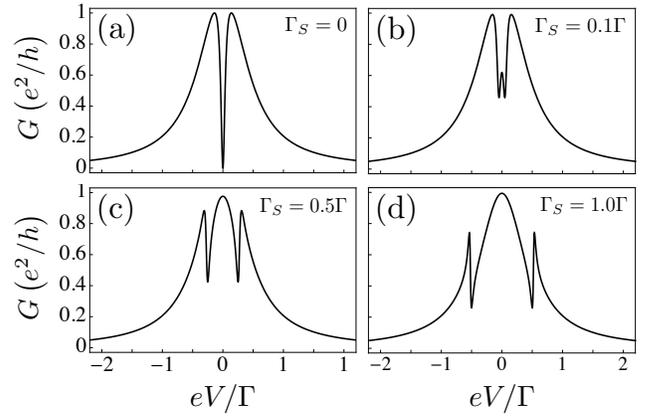}}
\caption{Conductance (in units of $\frac{e^{2}}{h}$) of the QD--QD--S system as a function of the bias voltage $eV/\Gamma$  for $t=0.14\Gamma$ and different values of $\Gamma_{S}$. 
We assume that the embedded QD is symmetrically coupled to the lead L and R and $\epsilon_{1}=\epsilon_{2}=0$.}
\label{fig1}
\end{figure}

\textit{Results and discussions}. Results of the transport properties at zero temperature ($T = 0$) are presented next. We assume a symmetric dot-lead couplings $\Gamma_{L}=\Gamma_{R}$, $\Gamma=\Gamma_{L}+\Gamma_{R}$ will be considered as the energy unit and $E_{F} = 0$. 

In Fig. \ref{fig1} we show the conductance (in units of $e^{2}/h$) in function of the bias voltage $eV/\Gamma$ for the QD--QD--S system. We can observe how the conductance changes for several values of the coupling with the superconducting lead, $\Gamma_{S}$. When the superconducting lead is not connected to the 2QDs system, (see Fig. \ref{fig1} (a)) we can see a symmetric Fano line shape that vanishes at $eV/\Gamma=0$, because a regular fermionic zero mode occurs on the end of the wire.
Once the superconductor is weakly coupled we can observe two small antirresonances. As the coupling with the superconducting lead ($\Gamma_{S}$) increases, the antirresonances become into two accentuated Fano antirresonances, whose minimal do not fall to zero, that is, the antirresonances have a complex Fano factor $q_{S}$. These Fano antirresonances originate from the interference between electrons traversing in different paths (a resonant and a non resonant one \cite{Fano}) when they propagate from the left to the right leads. The antirresonances, located around $\pm\Gamma_{S}/2$ have identical shape but an opposite sign of the parameter $q_{S}$ and are due to the Andreev Bound States (ABS). \cite{AMCalle} 

As mentioned in the paper of Zeng et. al. \cite{Zeng2017}, the Fano profile in the conductance can be used to detect the MBSs in the topological system QD--MBSs.
Authors further argue that the QD--MBSs can be mapped into a T-shape quantum dot system without a superconductor lead (denoted as QD--QD system), except when the threading magnetic flux through the ring is $\phi=\left(2n+1\right)\pi$. At this condition the QD--MBSs and the QD--QD are not equivalent to each other, since the conductance at Fano resonances in the QD--MBSs is not suppressed to zero, while the conductance at Fano resonances in the QD--QD does.\cite{Zeng2017} 
Motivated by those results, we realized an alternative non-topological QD--QD system that is fully equivalent to the topological QD--MBSs configuration. 

\begin{figure}[t!]
\centerline{\includegraphics[scale=0.48,angle=0]{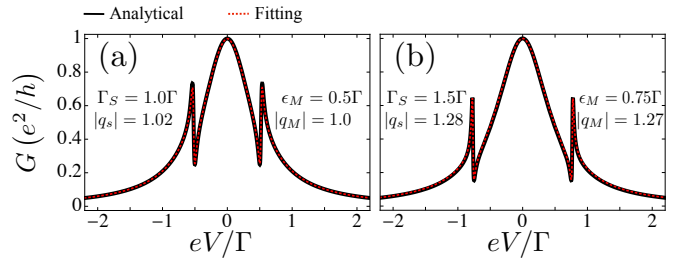}}
\caption{Conductance as a function of the bias voltage $eV/\Gamma$. For the QD--QD--S system we use the parameters $t=0.14$ and $\epsilon_{1}=\epsilon_{2}=0$. 
For the QD-MBSs system we use the magnetic flux phase $\phi=\pi$ and $|\lambda_{1}|=|\lambda_{2}|=0.1\Gamma$. 
The dotted (red) line corresponds to the fitting with the convolution of the Fano and the Breit Wigner lineshape.}
\label{fig2}
\end{figure}

As we display in Fig. \ref{fig1}, when a superconducting lead is connected, the transport properties exhibit two Fano resonances that do not drop to zero. In order to compare the QD--QD--S and the QD--MBSs to each other, we show in Fig. (\ref{fig2}) the conductance as a function of the bias voltage for both systems. In this figure, we illustrate the equivalence by showing the case where the conductance of the QD--QD--S and the QD--MBSs are identical to each other. Here, the black continuous curve describe the analytical conductance (calculated using (\ref{G211}) and (\ref{G_Zeng})), and the red dotted line describes the fitting (calculated from a convolution of Breit-Wigner and Fano function) as we describe later. The conductance for the QD--QD--S (QD--MBSs) system is shown for the superconductor (Majorana fermions) coupling $\Gamma_{S}=1\Gamma$ ($\epsilon_{M}=0.5\Gamma$) and $\Gamma_{S}=1.5\Gamma$ ($\epsilon_{M}=0.75\Gamma$). 
In this figure we observe two Fano resonances located approximately in $\pm\Gamma_{S}/2$ and $\pm\epsilon_{M}$ for the QD--QD--S and QD--MBSs systems, respectively. Notice that the modulus of the Fano factors $|q_s|$ and $|q_M|$  increase with increasing $\Gamma_{S}$ and $\epsilon_{M}$, respectively.
The Fano resonances in the QD--MBSs are originated from the interference between those electrons through the QD without going in the nanowire and those going into the nanowire.

In order to demonstrate the equivalence between the QD--QD--S and QD--MBSs to each other, we fit the the analytical conductance as $G_{X}\approx G_{F}\times G_{BW}$. $G_{X}$ is the convolution of a Fano line shape $G_{F}=\mathcal{A}_{X} \hspace{0.1cm}\frac{|\frac{\omega+\mathcal{B}_{X}}{\mathcal{C}_{X}}+q_{X}|^{2}}{\left(\frac{\omega+\mathcal{B}_{X}}{\mathcal{C}_{X}}\right)^{2}+1}$ and a Breit-Wigner line shape $G_{BW}=\frac{\mathcal{F}_{X}}{\omega^{2}+\left(\frac{\Lambda_{X}}{2}\right)^2}$ and the subindex $X$ is used to denote the QD--QD--S and QD--MBSs by $X=S$ and $X=M$, respectively.  
After a fitting procedure, these coefficients jointly with the parameters $\mathcal{A}_{X}$, $\mathcal{B}_{X}$, $\mathcal{C}_{X}$ and $\mathcal{F}_{X}$, $\Lambda_X$ are analytically obtained, which for the QD--QD--S system are given as 

\begin{eqnarray}
\label{eqsS}
\small
\nonumber
\mathcal{A}_{S}&=&\left(2\frac{\Gamma_{L}\Gamma_{R}}{\Gamma_{L}+\Gamma_{R}}\right)\frac{\Gamma}{2} \left(\frac{1}{\left(\frac{\Gamma}{2}\right)^{2}+\left(\frac{\Gamma_{S}}{2}\right)^{2}}\right), \\ \nonumber
\mathcal{B}_{S}&=&-\frac{8t^{4}\left(\Gamma^{2}-\Gamma^{2}_{S}\right)+4t^{2}\Gamma^{2}_{S}\left(\Gamma^{2}+\Gamma^{2}_{S}\right)+\Gamma^{2}_{S}\left(\Gamma^{2}+\Gamma^{2}_{S}\right)^{2}}{2\Gamma_{S}\left(\Gamma^{2}+\Gamma^{2}_{S}\right)^{2}}, \\  
\mathcal{C}_{S}&=&2t^{2}\Gamma\left(\frac{\Gamma^{2}+\Gamma^{2}_{S}-4t^{2}}{\left(\Gamma^{2}+\Gamma^{2}_{S}\right)^{2}}\right), 
\end{eqnarray}

\noindent whereas for the QD--MBSs system such parameters are  
\begin{eqnarray}
\label{eqsM}
\small
\nonumber
\mathcal{A}_{M}&=&\left(2\frac{\Gamma_{L}\Gamma_{R}}{\Gamma_{L}+\Gamma_{R}}\right)\frac{\Gamma}{2}\left(\frac{1}{\epsilon^{2}_{M}+\left(\frac{\Gamma}{2}\right)^{2}}\right), \\ \nonumber
\mathcal{B}_{M}&=&-\left(1+\frac{2|\lambda_{1}|^{2}}{\epsilon^{2}_{M}+\left(\frac{\Gamma}{2}\right)^{2}}\right)\epsilon_{M}, \\ 
\mathcal{C}_{M}&=&\frac{\Gamma \hspace{0.1cm} |\lambda_{1}|^{2}}{\left(\frac{\Gamma}{2}\right)^{2}+\epsilon^{2}_{M}},
\end{eqnarray}

\noindent and finally, for both systems the absolute value of $q_{X}$ for $X=S$ and $X=M$,  are given by

\begin{equation}
\label{qSC}
|q_{S}|= \sqrt{\frac{1}{2}+\frac{1}{2}\left(\frac{\Gamma_{S}}{\Gamma}\right)^{2}+\frac{4t^{2}}{\Gamma^{2}+\Gamma^{2}_{S}-4t^{2}}}, \hspace{0.1cm}
\end{equation}

\begin{equation}
\label{qMBS}
|q_{M}|= \sqrt{\frac{1}{2}+2\left(\frac{\epsilon_{M}}{\Gamma}\right)^{2}},
\end{equation}

\noindent and $\mathcal{F}_{X}=\frac{G_{\text{res}} \left(\omega_{\text{res}}\right)^2}{A \left| q_{X}\right| ^2-\frac{G_{\text{F0}}}{G_0}G_{\text{res}}}$, $\Lambda_{X} =4\frac{G_{F0}}{G_{0}} \hspace{0.1cm} \frac{G_{\text{res}} \left(\omega_{\text{res}}\right)^2}{A \left| q_{X}\right| ^2-\frac{G_{\text{F0}}}{G_0}G_{\text{res}}}$,  where $G_{\text{F0}}=G_F(\omega =0)=\frac{A_{X} \left(\left(B_{X}+C_{X} \hspace{0.1cm} \Re{q_{X}}\right){}^2+C_{X}^2 \hspace{0.1cm} \Im{q_{X}}^2\right)}{B_{X}^2+C_{X}^2}$, $G_0=G_{\text{analit}}(\omega =0)$, $G_{\text{res}}=G_{\text{analit}}\left(\omega =\omega_{\text{res}}\right)$ and $\omega_{\text{res}}=-\mathcal{B}_{X}$. $G_{\text{analit}}$ is calculated using Eqs. (\ref{condG}) and (\ref{G211}) for QD--QD--S, and Eq. (\ref{G_Zeng}) for QD--MBSs. 

\begin{figure}[t!]
\centerline{\includegraphics[scale=0.4,angle=0]{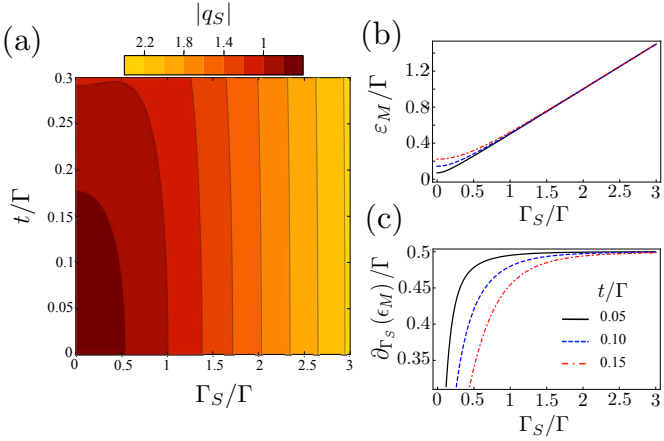}}
\caption{a) Contour plot of the Fano factor $|q_{S}|$ for the QD--QD--SC system as a function of the parameters $\Gamma_{S}/\Gamma$ and $t/\Gamma$. In (b) and (c) we display the linear and derivative relation between $\epsilon_{M}$ and $\Gamma_{S}$ for different values of $t$.
}
\label{fig3}
\end{figure}

The absolute value of the Fano factor $|q_{M}|$ increase as we increase the coupling $\epsilon_{M}$ between the Majorana fermions. Increasing (decreasing) $\epsilon_{M}$ also increases (decreases) the distance between the Fano antiresonances. When $\epsilon_{M}=0$ the two Fano resonances merge into one resonance, hence the two MBSs do not overlap, which in terms of conductance means that the peak value of the dot conductance at zero bias (when the dot is symmetrically coupled to the leads) is $e^{2}/2h$.\cite{Zeng2017}
At this point, it is relevant to mention that when the nanowire is in its topological phase, the zero temperature value of the dot conductance at $\epsilon_{M}=0$, is predicted to be $e^{2}/2h$. In contrast, when the wire is in its trivial phase, the conductance peak value is zero whenever a regular fermionic zero mode occurs on the wire ends \cite{Liu2011}, as occurs in Fig. \ref{fig1} (a) for $\Gamma_{S}=0$.

From Fig.(\ref{fig2}) we can observe that the QD--QD--S can be mapped to a QD--MBSs, when $\phi=\left(2n+1\right)\pi$; that is, when the nanowire is in its topological phase (the one with Majorana zero modes at the end of the nanowire) and when $\epsilon_{M}\neq 0$. This equivalence arises from the fact that in the QD--QD--S system the Fano resonances are due to the ABSs, which are particle-hole excitations of the SC whose field operators, written as $f^{\dag}=\left(\gamma_{1}+\dot{\imath}\gamma_{2}\right)/\sqrt{2}$ and $f=\left(\gamma_{1}-\dot{\imath}\gamma_{2}\right)/\sqrt{2}$, can be decomposed into a pair of Majorana operators $\gamma_{1}=\gamma^{\dag}_{1}$, $\gamma_{2}=\gamma^{\dag}_{2}$. In particular, low energy ABS can be viewed as a pair of overlapping MBSs \cite{Yoreg1}. 
The equivalence can also be explained from the QD--MBSs side. When the nanowire is in its topological non-trivial phase the Majorana zero modes appear at the ends of the nanowire, which can be represented by $\gamma_{1}=\left(f^{\dag}+f\right)/\sqrt{2}$ and $\gamma_{2}=\dot{\imath}\left(f^{\dag}-f\right)/\sqrt{2}$. Since these fields are the self-adjoint $\gamma_{1}=\gamma^{\dag}_{1}$, $\gamma_{2}=\gamma^{\dag}_{2}$, they represent mixtures of particle-hole states (ABS). Then, spatial overlaping Majoranas are indistinguishable from an ordinary ABS.
Therefore, we can conclude that the non-topological QD--QD--S system and the topological QD--MBSs system are equivalent.

In order to reinforce this statement,
in Fig. \ref{fig3} (a) we show the contour plot of the Fano factor $|q_{S}|$ as a function of $t/\Gamma$ and $\Gamma_{S}/\Gamma$. When $\Gamma_{s}\gg t$ we observe that the dependence of $|q_{S}|$ with the parameter $t$ is irrelevant. We also observe a monotonous growth of $|q_{S}|$ with $\Gamma_{S}$ for any value of $t$. On the other hand, when the previous condition is not fulfilled (when $\Gamma_{S}\lesssim t$ or $\Gamma_{S}\approx t$), the contour plot indicates that the dependence of $|q_{S}|$ with the parameters $t$ and $\Gamma_{S}$ are equally important and in consequence the growth of $|q_{S}|$ with $\Gamma_{S}$ is not monotonous as for the case $\Gamma_{S}\gg t$. However, in order to underpin our previous analysis about the equivalence between both systems as shown in Fig. \ref{fig2},
in this paper we focus our attention to the limit $\Gamma_{S}\gg t$. Indeed, an one-to-one relation between $\epsilon_{M}$ and  $\Gamma_{S}$ can be obtained for several values of the parameter $t$. For this, we assumed that $|q_s|=|q_M|$ then from equations (\ref{qSC}) and (\ref{qMBS}) we obtain that
\begin{equation}
\label{EGamma}
\epsilon_{M}= \sqrt{\left(\frac{\Gamma_{S}}{2\Gamma}\right)^{2}+\frac{4t^{2}}{\Gamma^{2}+\Gamma^{2}_{S}-4t^{2}}}.
\end{equation}
From this equation we observe a linear relation between $\epsilon_{M}$ and $\Gamma_{S}$ with a proportional term of $1/2$ at the limit $\Gamma_s\gg t$.
This can be verified in Fig. \ref{fig3} (b) and Fig. \ref{fig3} (c) where the linear and derivative relation between $\epsilon_{M}$ and $\Gamma_{S}$ is displayed. It is worth noticing here that the assumption $|q_S| =|q_M|$ is more straight at our limit of interest. Indeed, by taking the limit $t\ll \Gamma _s$ in the equation (\ref{qSC}) we obtain $|q_{S}|\left(t\ll \Gamma _s\right)\approx \sqrt{\frac{1}{2}+2\left(\frac{\Gamma_{S}}{2\Gamma}\right)^{2}}$,  which coincide in an identical mathematical aspect with the results obtained for $|q_{M}|$ in equation (\ref{qMBS}) as far as we map $\epsilon_{M}$ and $\Gamma_{S}/2$ to each other. 
Analogously, in the limit $t\ll \Gamma _S$, we obtain $\mathcal{B}_{S}\approx-\left(1+\frac{t^{2}}{\left(\frac{\Gamma_{S}}{2}\right)^{2}+\left(\frac{\Gamma}{2}\right)^{2}}\right)\frac{\Gamma_{S}}{2}$ and $\mathcal{C}_S\approx \frac{\Gamma \hspace{0.1cm} \frac{t^2}{2}}{\left(\frac{\Gamma_{S}}{2}\right)^2+\left(\frac{\Gamma}{2}\right){}^2}$, which are respectively identical to the quantities $\mathcal{B}_{M}$ and $\mathcal{C}_{M}$ with the mapping $\Gamma_S/2\rightarrow\epsilon_{M}$ and $t^2\rightarrow 2|\lambda_1|^2$. To summarize, we can observe that coefficients $\mathcal{A}_{S}$, $\mathcal{B}_{S}$, $\mathcal{C}_{S}$ and $q_{S}$ (Eqs. (\ref{eqsS}) and (\ref{qSC})) are identical to $\mathcal{A}_{M}$, $\mathcal{B}_{M}$, $\mathcal{C}_{M}$ and $q_{M}$ (Eqs. (\ref{eqsM}) and (\ref{qMBS})), respectively, as far as we restrict to the limit $\Gamma_S\gg t$ and we map  $\Gamma_S/2\rightarrow\epsilon_{M}$ and $t^2\rightarrow 2|\lambda_1|^2$. Note that under these conditions the parameters $\Lambda_X$ and $\mathcal{F}_X$ are identical for both systems. Accordingly, we can conclude that $\Gamma_S/2$ and $\epsilon_{M}$ play identical roles in their corresponding system, therefore showing that the QD--QD--S and QD--MBS are equivalent to each other from their transport properties point of view.

Finally, it is worth to mentioning that a complex Fano factor $q$ is an indication of the broken time reversal symmetry, \cite{Kobayashi_Fano_2003} which can be introduced, for instance, with the application of a magnetic field. This is the case of the QD--MBSs where the Fano antirresonances do not fall to zero (as shown in Fig. \ref{fig2}. (c) and (d)) due to the applied magnetic field. 
This notable feature can be used to detect the Majorana zero energy modes in the system. 
In contrast, in the QD--QD--S we observe the same non-zero Fano antirresonances as shown in Fig. \ref{fig2}. (a) and (b). This is a non-trivial results worth of being highlighted given that the ground state of a superconductor does not conserve the number of particles and maintains the time reversal symmetry.
One possible explanation to this behaviour could be as follows. Despite of the idea of Andreev reflected hole as the time reverse of the incident electron, we can realize that this scheme breaks down \cite{Why?} because
there is a phase shift acquired upon Andreev reflections that spoils the time-reversing properties; 
therefore we could deduce that the superconducting lead has the role of introducing this phase and consequently, breaking the time-reversal symmetry in the system QD--QD--S.

\textit{Summary}. We demonstrated an interesting case where the electronic transport properties of MBSs in a topological superconductor system (QD--MBSs) and a non-topological T-shaped quantum dot structure (QD--QD--S) are identical to each other. We demonstrated that the Fano factor $q_{S}$ of the QD--QD--S is equivalent to the Fano factor $q_{M}$ of the QD-MBSs. Furthermore, we found that the non-zero value of the Fano (anti)resonance in the conductance, which is an evidence of MBSs, \cite{Zeng2017} is due to a complex Fano factor $q_{M}$. Analogously, the complex Fano factor $q_{S}$ is a signal of the fact that the superconducting lead is introducing a phase in the QD--QD--S. We argue that the equivalence between both systems is due to this phase. We believe that our results can stimulate further works toward the understanding of transport properties of topological quantum hybrid structures from conventional non-topological quantum systems. 


\begin{acknowledgments}
A.M.C  gratefully acknowledge I. C. Fulga for fruitful discussions.
\end{acknowledgments}

\section*{References}
\bibliographystyle{apsrev4-1}

\end{document}